\documentclass[3p,times]{elsarticle}

\usepackage{lineno,hyperref,amssymb}
\modulolinenumbers[5]

\journal{Comptes Rendus Physique}

\def\be{\begin{equation}}
\def\ee{\end{equation}}     
\def\bfi{\begin{figure}}
\def\efi{\end{figure}}
\def\bea{\begin{eqnarray}}
\def\eea{\end{eqnarray}}

\begin{document}

\begin{frontmatter}

\title{Coarsening in inhomogeneous systems}

\author{Federico Corberi}
\address{Dipartimento di Fisica ``E.~R. Caianiello'', and INFN, Gruppo Collegato di Salerno, and CNISM, Unit\`a di Salerno,Universit\`a  di Salerno, 
via Giovanni Paolo II 132, 84084 Fisciano (SA), Italy.}
\ead{corberi@sa.infn.it}

\begin{abstract}
This Article is a brief review of coarsening phenomena occurring in systems where
quenched features - such as random field, varying coupling constants or lattice vacancies -
spoil homogeneity. We discuss the current understanding of the problem 
in ferromagnetic systems with a non-conserved scalar order parameter 
by focusing primarily on the form of the growth-law of the ordered domains
and on the scaling properties. 
\end{abstract}

\begin{keyword}
\texttt{Coarsening, Scaling, Disorder}
\end{keyword}

\end{frontmatter}


\section{Introduction} \label{intro}

Coarsening can exhibit rather different features
in systems where homogeneity holds or not. 
In order to
keep the discussion at the simplest level I will
focus on the class of ferromagnetic systems with a scalar order-parameter (spin),
such as those whose properties can be described by
the Ising model or, in a continuum approach, by the usual Ginzburg-Landau
free energy. 
Inhomogeneous systems with a vectorial order-parameter 
are much less studied and will not be discussed here.
Besides, the attention will be restricted to a dynamics
that does not conserve the order-parameter.
This descriptions are in principle
suited, e.g., for a magnetic solid. 
The aim of the article is not
to provide a comprehensive description of the behavior of 
existing real systems, but rather to discuss at the simplest possible level
the new features introduced by the presence 
of inhomogeneities.

\begin{figure}[h]
  \vspace{1cm}
    \centering
\rotatebox{0}{\resizebox{.4\textwidth}{!}{\includegraphics{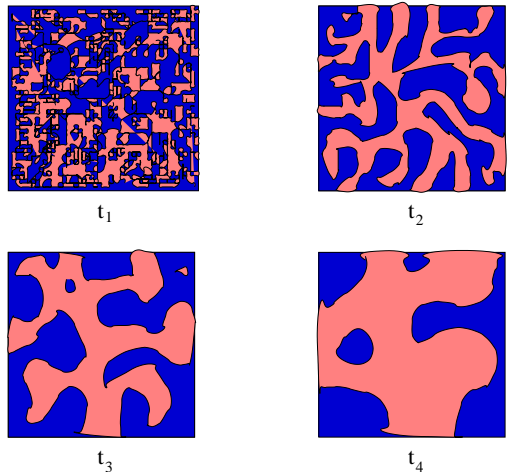}}}
\caption{A pictorial representation of a coarsening process.
Blue and pink regions are the growing domains of the two low-temperature
equilibrium ordered phases. The four panels correspond to snapshots
of the system at subsequent times $t_1<t_2<t_3<t_4$.}
\label{domains}
\end{figure}

In the kind of systems we are considering coarsening is usually
observed after a quench from an high-temperature equilibrium 
configuration to a temperature below the critical one.
Relaxation toward the new equilibrium state then occurs by the formation
and growth of domains inside which the system is basically
equilibrated in one of the two symmetry-related low-temperature
equilibrium phases (see Article by L.F. Cugliandolo in this Volume). 
For a magnet, these are the two possible
ordered configurations with positive or negative magnetization $\pm m$.
The non-equilibrium behavior occurs on the interfaces which move
in order to increase the typical domain's size.
A pictorial representation of the process is depicted in
figure \ref{domains}.

The distinguishing characteristic of coarsening in
homogeneous systems is the scaling symmetry \cite{purescaling}. 
Accordingly, in the late stage of
the process, configurations of the system at different times
are statistically equivalent if lengths are measured in units of
the characteristic size $L(t)$ of the domains, which grows 
algebraically like 
\be
L(t)=a t^{1/z},
\label{ldit}
\ee
where $a$ is a constant, and $z=2$ in the case without order-parameter 
conservation considered here \cite{purescaling}. 
The scaling symmetry is mirrored by the form taken by different
quantities, as for instance correlation functions.
Using the language of spin systems
the two-time spin-spin correlation 
$G(r,t)=\langle S_i(t)S_j(t_w)\rangle$, where $S_i=\pm 1$ are spin variables
on sites $i$ and $j$ of a regular lattice  
at a distance $r$, obeys 
\be
G(r,t_w,t)=g\left [\frac{r}{L(t_w)},\frac{L(t)}{L(t_w)}\right ],
\label{scalg}
\ee
where $g(x,y)$ [with $x=r/L(t_w)$ and $y=L(t)/L(t_w)$) 
is a scaling function (and, similarly, $s(x)$ and $c(y)$ below].
The form (\ref{scalg})  contains, as special cases, the scaling 
$S(r,t)=s(x)$ of 
the equal-time correlation function $S(r,t)=G(r,t,t)$ and that
$C(t,t_w)=c(y)$ of the autocorrelation function $C(t,t_w)=G(0,t,t_w)$.

Eq. (\ref{scalg}) is, in principle, strictly obeyed only 
in a quench to a vanishing final temperature $T_f=0$. The effect of a finite
quench temperature is to modify the behavior (\ref{scalg}) of correlations
at small distances [i.e. $r<\xi (T_f)$, where $\xi (T_f)$ is the 
coherence length of the equilibrium state at $T_f$ (which is 
small except in the critical region)] and small time-differences
$t-t_w<\xi (T_f)^z$]. For simplicity, in the following our discussion about
scaling properties will always be restricted to sufficiently large 
space and time scales where such corrections can be discarded.
A thorough discussion of the whole scaling form taken by
$G$ at finite $T_f$ is contained in \cite{tuttoscal}.

With the possible exception of liquid systems, whose comprehension
however turns out to be more difficult (see Article by S. Das, S. Roy and J. Midya 
in this Volume), real systems are almost never homogeneous,
due to the presence of quenched lattice defects or because of external
disturbances as site-dependent applied fields. 
Experiments show that these sources of inhomogeneities deeply perturb
the coarsening process and the scaling properties. Indeed, although 
some kind of scaling symmetry comparable to Eq. (\ref{scalg}) 
is sometimes reported \cite{exprb3} for the correlation functions, 
the asymptotic growth-law is never observed in the form (\ref{ldit})
and, instead, a much slower increase of the ordered patterns,
usually of a logarithmic type \cite{exprf,exprb,exprb2,exprb3}), occurs.  

In this Article we will mainly review 
how and why the growth-law and the dynamical scaling symmetry,
expressed  by Eqs. (\ref{ldit},\ref{scalg}), are modified 
in an inhomogeneous coarsening system.

\section{Models} \label{mods}

\subsection{Homogeneous systems} \label{modhom}

As an optimal playground to study coarsening 
I consider the description of an homogeneous system 
in terms of the Ising model with Hamiltonian
\be
{\cal H}(\{S_i\})=-J\sum _{<ij>}S_iS_j,
\label{ham}
\ee
where $\{S_i\}$ is a configuration of $N$ spin variables $S_i=\pm 1$ 
and $i=1,\dots,N$ are sites of a regular
$d$-dimensional lattice. By {\it homogeneous} it is meant that
the Hamiltonian [for instance the one (\ref{ham})] does not select
any particular spacial position and all places are {\it a priori} equal.
Alternatively one can say that, although a single realization of the system
can be inhomogeneous (i.e. due to the presence of domains)
local observables do not depend on space after the thermal averaging.

A dynamics is introduced
by means of a master-equation with transition rates $w(\{S_i\}\to \{S'_i\})$
regulating the hopping between configurations. In the following
detailed balance with respect to the equilibrium Boltzmann-Gibbs
measure $e^{-\beta {\cal H}}$ will be assumed for the $w$'s, implying that 
the equilibrium state is spontaneously approached. 
I will also assume, again for simplicity,
that the sequential evolution of single spins occurs and that
the order parameter is not conserved. 

For such a model, it is well known that, after a quench from an high-temperature
equilibrium ensemble to a temperature $T_f=(k_B\beta_f)^{-1}$ 
below the critical one $T_c$, 
dynamical scaling holds together with Eqs. (\ref{ldit},\ref{scalg})
for any $T_f<T_c$ including $T_f=0$.
 
In the following I will discuss how the most common sources of inhomogeneities, 
i.e. varying coupling constants, external fields and lattice defects can be included in
the above model. I will restrict
the discussion to the case in which such features are quenched, namely 
their position, strength etc. is not thermalized and do not vary appreciably
during the coarsening phenomenon. 

\subsection{Varying coupling constants} \label{RBIM}

If the coupling constants are not uniform the Ising Hamiltonian
reads
\be
{\cal H}(\{S_i\})=-\sum _{<ij>}J_{ij}S_iS_j.
\label{hamrb}
\ee
where $J_{ij}=J_0+\theta _{ij}$ and the 
$\theta $'s are usually assumed to be uncorrelated random numbers
with $\overline {\theta_{ij}\theta_{k\ell}}={\cal J}^2 \delta _{i,k}\delta_{k,\ell}$.
Here and in the following an over-bar denotes an average taken
over the realization of the disorder. 
The probability distribution of the $\theta$'s must be such that $J_{ij}\ge 0$ 
(otherwise frustration effects could be present which may radically
spoil the ferromagnetic character of the system with a drastic
modification of the equilibrium and non-equilibrium properties).
This is known as the random-bond Ising model (RBIM).
Notice that, since the distribution of the coupling constant is usually
assumed to be symmetric around $J_0$, in this model there is
a maximum value $J_{ij}=J_{max}\le 2J_0$ of the strength of the bonds.

\subsection{External fields}

Under the action of  a site-dependent external magnetic field $h_i$ 
the Ising Hamiltonian reads
\be
{\cal H}(\{S_i\})=-J\sum _{<ij>}S_iS_j-\sum _i h_i S_i.
\label{hamrf}
\ee
A useful model is the one where 
$h_i$ is an uncorrelated random variable with expectations 
$\overline {h_i h_j}=h^2\delta _{ij}$, extracted from a given distribution
(the simplest case is a symmetric bimodal distribution $h_i=\pm h$).
With these specifications one has the so-called random-field Ising model
(RFIM).

\subsection{Lattice vacancies} \label{latvac}

The effect of quenched vacancies can be taken into account 
by defining a dilution variable $\chi _{ij}=0,1$ on the lattice
in such a way that the Hamiltonian reads
\be
{\cal H}(\{S_i\})=-J\sum _{<ij>}\chi_{ij}S_iS_j.
\label{hamdil}
\ee
If the dilution variable is factorized as $\chi _{ij}=\chi _i \chi_j$ 
the system has site-dilution and the model is usually referred to as
the site-diluted Ising model (SDIM). Another possibility is to have 
bond-dilution: In this case $\chi _{ij}$ is defined on the couples of
neighboring spins.

In many cases the $\chi$'s can be considered random numbers,
usually uncorrelated in space, such that a fraction $D$ of the
lattice sites - or bonds - are diluted, namely they are associated
to $\chi _i=0$ or $\chi _{ij}=0$ respectively. 
This models are usually denoted as diluted Ising
models. 

For dilutions larger than the limiting value $D_c=1-P_c$ -- where $P_c$ is 
the critical percolation probability --  a spanning network does not exist. This case is not interesting to us
because coarsening implies spins to live on a connected graph 
extending throughout the system.
In particular, since $D_c=0$ in one dimension,
these models will not be considered in $d=1$. 

For the description of other systems, like
those where spins are defined on the edges of a fractal network,
which can be either deterministic or random, 
the $\chi$'s are taken in
order to reproduce the observed topology of such a network.
I will briefly discuss this issue later.

Of course, possible modifications of the models introduced so far - 
i.e. those where the random variables are correlated or where
more than one source of disorder is present - are possible,
but I will not considered them here.

\section{Downhill versus activated coarsening} \label{downvsact}

In an homogeneous system described by the model of Sec. \ref{modhom},
coarsening occurs at any temperature including $T_f=0$.
This implies that the dynamics, whose main mechanism is a displacement 
of the interfaces which is regulated by their curvature 
(see article by A.A. Nepomnyashchy in this Volume), 
proceeds without thermal activation.

Quenched disorder or any other source of inhomogeneity in the system,
instead, usually introduces preferred positions where interfaces 
get pinned in local energy minima. The dynamics can then proceed only
by means of thermal activation. 

A simple argument shows how this
can slow down the kinetics with respect to the homogeneous case where 
Eq. (\ref{ldit}) holds. In view of the dynamical scaling property,
the typical pinning energy barrier $\Delta E$ at time $t$ is expected to 
depend on the configuration through $L(t)$ alone, $\Delta E=
f[L(t),K]$, where $K$ specifies the set of 
model parameters, as for instance $K=(J/T,h/T)$ in the RFIM or 
$K=(J/T,d)$ in the SDIM. In the following, among these various parameters
we will explicitly write only the one, denoted by 
$\epsilon$, which denotes the strength of the disorder
(as $h$ in the RFIM, or ${\cal J}$ in the RBIM).

The evolution is 
slowed down due to the Arrhenius time
$t_{esc}(\Delta E)\propto e^{\beta \Delta E}$ needed
to escape pinning barriers.
Making the simplifying assumption that this effect can be taken into account by a simple
rescaling of time ($t\to t/t_{esc}$) in Eq. (\ref{ldit}) one has
\be
L(t,\epsilon)=a t^{1/2}e^{-\frac{\beta}{2}\Delta E}
\label{genl}
\ee
(In the presence of quenched disorder, quantities as
$L(t)$ or $G(r,t,t_w)$ are defined throughout as 
averaged both over the thermal history and the disorder realizations).
The next point is to establish the form of $f[L,\epsilon]$.
Although this cannot be done in general, most of the systems
can be divided into three classes \cite{maz}  
according to the way $\Delta E$ depends
asymptotically on $L(t)$, namely if i) $\Delta E$ approaches a constant
value or ii) it diverges algebraically or iii) logarithmically with $L(t)$.
Let us detail below these cases:

\begin{itemize}
\item{i) $\lim _{L\to \infty}f[L,\epsilon]\le c(\epsilon)$, 
where $c(\epsilon)$ is usually an increasing function
of disorder (with $c=1$ for $\epsilon =0$), 
but it is constant with respect to $L$.

This implies that there is an upper limit
to the height of barriers. Using the fact that barriers are
approximately constant in Eq. (\ref{genl}) one finds the same growth-law 
of an homogeneous system but with a smaller pre-factor, of order
$ae^{-\frac{\beta}{2}c(\epsilon)}$,
which depends strongly on temperature.}

\item{ii) $\lim _{L\to \infty} f[L,\epsilon]=b^{-1}(\epsilon)L^\psi$, where $\psi >0 $ is
an exponent and $b(\epsilon)$ is another constant.
In this case barriers grow algebraically with the domain's
size. This leads to a logarithmic growth 
\be
L(t)\simeq [b\beta ^{-1}\ln t]^{1/\psi}
\label{loglaw}
\ee
for large times.}

\item{iii) $\lim _{L\to \infty}f[L,\epsilon]=z(\epsilon)\ln L$, with 
$z(\epsilon)$ a constant. Plugging into Eq. (\ref{genl}) one finds
the large time behavior 
\be
L(t)\propto t^{1/\zeta},
\label{plawT} 
\ee
with an exponent $\zeta = 2+\beta z$ which depends on temperature and, possibly,
on other system's parameters.}

\end{itemize}

\subsection{Role of the lower critical dimension}

The models introduced above [including the usual Ising model (\ref{ham})]
have a finite critical temperature only in spatial dimension
larger than the lower critical one $d_L$.
If $d\le d_L$ a non-interrupted coarsening process can be observed
only when quenching to $T_f=0$ (in an infinite system), because for
any $T_f>0$ the ordered phase is not sustained.

Despite this, by quenching to a very low
temperature $T_f$, if the dynamics does not require activation 
one usually does observes coarsening up to a certain
time $t_{eq}(T_f)$ such that $L(t_{eq})\simeq \xi (T_f)$, where 
$\xi (T_f)$ is the final equilibrium coherence length.
For instance, this is what happens in the usual Ising model
in $d=d_L=1$ where $\xi (T_f)\simeq e^{2\beta _f J}$.
At low temperatures $t_{eq}(T_f)$ grows very large,
and for $t\ll t_{eq}(T_f)$ the kinetics is analogous to the one
at $T_f=0$. 

When homogeneity is spoiled the dynamics right at $T=0$ is frozen 
due to the energetic barriers discussed above which pin the system in 
metastable states. 
The way out is to quench to a temperature high enough to drive activated 
coarsening, but low enough
to inhibit the nucleation of equilibrium fluctuations. As for the
clean case, as long as $L(t)$ is
smaller than the equilibrium correlation length $\xi (T_f)$, 
one observes the same coarsening behavior as in the $T_f=0$ quench.

These considerations apply to all the one-dimensional cases that will be discussed in Section \ref{coars1d}
as well to other systems, that will be considered
in the following, for which $T_c=0$, as for instance the $d=2$ RFIM 
or some models defined on fractal lattices.

\subsection{Crossover structure} \label{cross}

The growth-laws i),ii),iii) of Section \ref{downvsact} 
are asymptotic behaviors of $L(t)$.
Usually these laws are expected when $L(t)$ exceeds
a certain characteristic length $\lambda (\epsilon)$ associated
to the presence of the disorder (or of other sources on inhomogeneities).
In order to understand the meaning of $\lambda (\epsilon)$ by means of a
specific example, let us consider the SDIM of Sec. \ref{latvac}.
For $D$ sufficiently small, vacancies are isolated 
with a typical inter-distance 
$\lambda (D)=D^{-1/d}$. As long as
$L(t)\ll \lambda (D)$ the domains do not feel the presence of
the dilution and the system behaves as an homogeneous one
(this will be further discussed in Section \ref{sdim2}).
Hence, whatever the
asymptotic behavior of this model is 
[either in class i), ii) or iii) of the previous
Section \ref{downvsact}], this can only show-up for sufficiently
large times such that $L(t)\gg \lambda (D)$.    
One can show that the crossover phenomenon at $L(t)\simeq \lambda (\epsilon)$, 
where the behavior of the system changes,
is a common feature, not restricted to the previous example of the SDIM,
as we will discuss further.

On the other hand, the behavior of the systems in the pre-asymptotic 
regime $L(t)\ll \lambda(\epsilon)$ is generally model-dependent: There 
are cases where, as in the SDIM discussed above or in models with a 
fraction of spins 
frozen in a random configuration \cite{grsr}, 
inhomogeneities are initially ineffective and one recovers the 
growth-law (\ref{ldit}) of the clean case, and others where 
they produce a different behavior.

As an example of this different behavior, let us consider
the RBIM. In this case there are random couplings on {\it any} bond, 
and -- if $T_f$ is sufficiently low -- this influences the process already at
the level of the fast process of single spin flips. 
Due to this, in the pre-asymptotic
regime $L(t)\ll \lambda ({\cal J})$ the disordered bonds alter 
the behavior of the 
system with respect to the clean case, thus spoiling Eq. (\ref{ldit}),
at variance with the SDIM [this will be further discussed in 
Sections \ref{rbim1},\ref{rbim2}].  
However, although determined by the disorder, this early behavior  
is not the asymptotic one, which sets in only when $L(t)$ is sufficiently
large to allow for the simple scaling argument of the previous section
\ref{downvsact} to become valid. Translating {\it sufficiently
large} in the condition $L(t)\gg \lambda ({\cal J})$ provides a 
definition of $\lambda ({\cal J})$ in this case (although physically
not as transparent as the one above for the SDIM).

I conclude this section by mentioning that the slowing down of the growth-law
discussed above is not restricted to those systems that -- when homogeneous --
support a non-activated dynamics. The simplest example is provided by  
systems with a conserved order-parameter. In the homogeneous case coarsening  proceeds 
asymptotically by means of the activated evaporation-condensation process. 
However, the presence of disorder \cite{rao2} or other inhomogeneities usually introduces
further energetic barriers slowing even more the kinetics.

\section{Scaling form}

Once the existence of a crossover length $\lambda (\epsilon)$ next to the
ordering length $L(t)$ is established, the next step is to ascertain
if a scaling symmetry is induced by the presence of such lengths
and, in case, how the form (\ref{scalg}) can be generalized.

A straightforward generalization of the 
form (\ref{scalg}) keeping into account the crossover length is the following 
\be
G(r,t_w,t,\epsilon)={\cal G}\left [\frac{r}{L(t_w)},\frac{L(t)}{L(t_w)},\frac{\lambda (\epsilon)}{L(t_w)}\right ],
\label{scalgd}
\ee
and similarly for other observable quantities. This amounts to the usual scaling
prescription: To each parameter entering $G$, here $r,t_w,t,\epsilon$, 
a characteristic length is associated, 
$r,L(t_w),L(t),\lambda (\epsilon)$ respectively, and these quantities
enters as adimensional ratios in a scaling function ${\cal G}$.

A debated question regards the dependence of the scaling functions,
like ${\cal G}(x,y,z)$, on the strength of the noise,
which occurs through the third entry $z=\lambda (\epsilon)/L(t_w)$.
According to an hypothesis, the so-called 
{\it superuniversality} \cite{superuniv},
the effect of disorder can be fully accounted for by the slower
growth of $L(t)$. This implies that disorder does not alter
all other properties of the coarsening process, among which -- for instance --
the geometry of the growing pattern, which is encoded in the form of 
${\cal G}$ in Eq. (\ref{scalgd}).
This would mean that no extra
dependence on $\epsilon $ is left in the scaling functions and,
for the correlations considered above, one should have
$G(r,t_w,t,\epsilon)={\cal G}\left [\frac{r}{L(t_w)},\frac{L(t)}{L(t_w)},1 \right ]=
g\left [\frac{r}{L(t,\epsilon)},\frac{L(t)}{L(t_w)}\right ]$, 
where $g(x)$ is the same as that in (\ref{scalg}) for the clean system.

Presently the superuniversality hypotheses has received confirmation
in some specific cases \cite{rflog3,otherfavorSU1,otherfavorSU2,SUfromdomains,lour,gyure}, but it has also been shown not to be obeyed in
other system \cite{LipZan10,Fisher,Decandia02,noirb,noirf}, as we will discuss below.  

\section{Coarsening in one dimension} \label{coars1d}

The effect of inhomogeneities on the coarsening process 
can be more easily understood in one-dimensional systems. In this case
the task is simplified by the fact that interfaces are point-like. 

\subsection{RBIM} \label{rbim1}

\begin{figure}[h]
  \vspace{1cm}
    \centering
\rotatebox{0}{\resizebox{.46\textwidth}{!}{\includegraphics{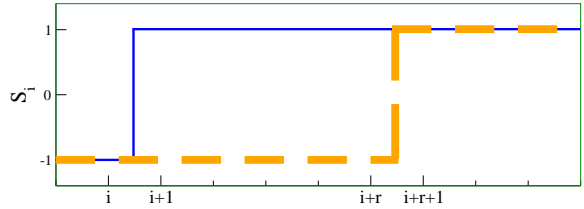}}}
\caption{An interface initially located between sites $i$ and $i+1$
(plotted in a continuous blue line) moves to a new position 
$r$ sites away (dashed orange line).}
\label{figbar}
\end{figure}

Let us suppose to have a single interface in the system 
originally located between sites 
$i$ and $i+1$ (namely $S_iS_{i+1}=-1$), as pictorially sketched
in figure \ref{figbar} (continuous-blue line).

\begin{figure}[h]
  \vspace{1cm}
    \centering
\rotatebox{0}{\resizebox{.65\textwidth}{!}{\includegraphics{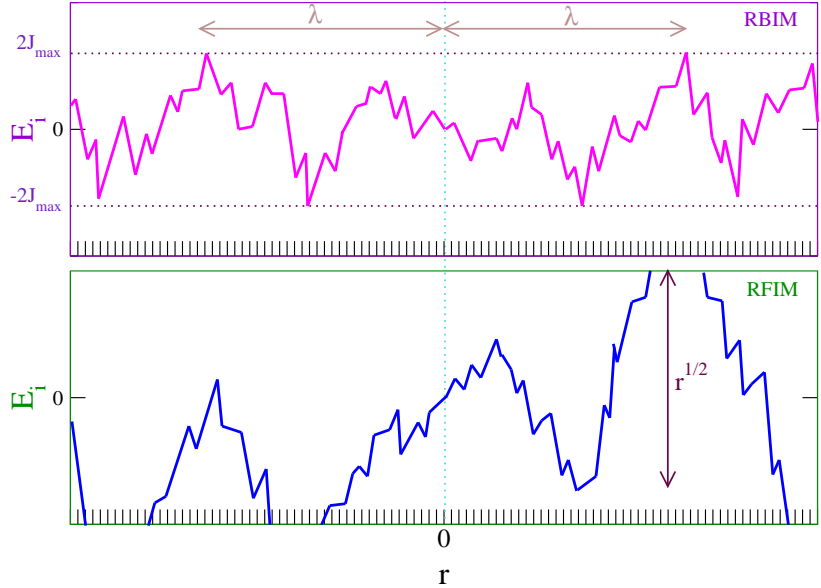}}}
\caption{The energy of a single interface is sketched against its
displacement $r$ for the RBIM (upper panel) and the RFIM (lower panel)
in $d=1$ (we set $E=0$ for $r=0$). 
For the RBIM the distance $\lambda $ where the maximum
energy bafrrier $4J_{max}$ is felt is also indicated. For the RFIM
the typical energy barrier increases as $r^{1/2}$.}
\label{figlandsc}
\end{figure}

Let us suppose that at a later time the interface moves a distance
$r$ away from the original position, namely between sites $i+r$ and
$i+r+1$ (dashed orange line in figure \ref{figbar}).
The energy change in this process is $\Delta E _{i,i+r}=2[J_{i+r,i+r+1}-J_{i,i+1}]$.
Since the coupling constants are random numbers this form of 
$\Delta E_{i,i+r}$ induces local minima 
and maxima in the potential landscape felt by an interface, namely the
energy cost of an interface as a function of its position, which is
sketched in figure \ref{figlandsc}. 

Recalling the discussion at the end of Sec. \ref{RBIM} there exists 
a maximum value $c(\epsilon)=4J_{max}$ for the height of the barriers and hence
the RBIM is an example of class i) of Sec. \ref{downvsact}. Therefore
Eq. (\ref{ldit}) holds for long times (but with $t<t_{eq}$) at low $T_f$. 

Clearly, interfaces must travel a certain distance -- or, equivalently, domains must
grow up to a certain crossover size $L(t)=\lambda $ -- in order to sweep a region containing
the barrier of maximal height $c$. 
According to the argument of Sec. \ref{cross},
pinning is effective from the very beginning of the process but
the difference with the asymptotic stage is that,
for $L(t)<\lambda $, $\Delta E$ may depend on $L(t)$. 
Accordingly the growth-law of the homogeneous system (\ref{ldit})
is not observed pre-asymptotically and, instead, one finds 
a ${\cal J}/T$-dependent power-law,
as in Eq. (\ref{plawT}) \cite{LipZan10}.

Regarding the superuniversality hypotheses it has been convincingly shown
not to hold in \cite{LipZan10}.

\subsection{RFIM}

In this case one has
\be
\Delta E _{i,i+r}=-\sum _{j=i}^{i+r}h_j\sim hr^{1/2},
\label{ensin}
\ee
where in the last
passage we have assumed a bimodal distribution $h_i=\pm h$
of the random field (but other possible choices do not make
significative differences) and we have used the central limit theorem.
Also in this case there are
local minima and maxima but the potential landscape where the point defect
moves increases with $r$, as it is pictorially shown in figure \ref{figlandsc}. 
Indeed, as Eq. (\ref{ensin}) shows,
the energy of the interface which has moved to a site at distance $r$ 
can be though itself as the position of a Brownian walker after 
$r$ steps: the energy landscape is therefore   
of the Sinai type \cite{Sinai}. As a consequence, the pinning energy
is not bounded as in the RBIM and increases algebraically with the typical 
distance $x(t)$ traveled by the interface, as indicated by Eq. (\ref{ensin}). 
Since it can be shown that also here,
as in the clean case, the average size of domains $L(t)$ grows proportional
to $x(t)$ \cite{Fisher}, one concludes that the $1d$ RFIM belongs to class 
class ii) of Sec. \ref{downvsact}. One indeed has $L(t)\propto (\ln t)^2$
in the large time domain \cite{Fisher, Decandia02}. 

Notice that monotonously increasing barriers means that they
are bound to prevail over the thermal energy scale $k_BT_f$ if 
$L(t)$ is large enough, namely for $L(t)\gg \lambda $, where $\lambda (h/T_f)\sim (h/T_f)^{-2}$
is the crossover length \cite{Decandia02}
that can be obtained balancing the two energetic contributions.

At early times $L(t)\ll \lambda$ the Sinai potential is washed out by thermal
fluctuations and one recovers the growth-law (\ref{ldit}) characteristic of the non-disordered
system. The $1d$ RFIM is therefore another example of a crossover with an
initial homogeneous-like behavior.

Also in this model superuniversality is
not obeyed \cite{Fisher,Decandia02}.

\section{Coarsening on fractal substrates} \label{fractals}

Coarsening occurring on fractal substrates that can be embedded 
in an Euclidean space
can be still described by a model with dilution, like the ones 
of Section \ref{latvac} [see Eq. (\ref{hamdil})], 
with suitable prescriptions for the $\chi 's$. 
In the following I will focus on the case with site-dilution. 
The problem has been
studied numerically \cite{noibarriers,noifract} both for 
spins defined on random fractal networks, like the
percolation cluster, and for deterministic ones, such as the Sierpinski fractal
and others.
These deterministic structures are built recursively upon replicating a native geometry
- e.g. a triangle for the Sierpinski gasket - of a certain size $\ell$.

After the quench, the typical signatures of the coarsening phenomenon are observed,
with the growth of an ordering-length and dynamical scaling.
Correlation functions can be cast in scaling
form, as in Eq. (\ref{scalg}): The scaling function is different
from the homogeneous case and depend on the fractal considered. 
Regarding the growth-law, one observes either a logarithmic increase of $L(t)$,
as for a system in class ii) [Sec. \ref{downvsact}], or a temperature-dependent
power growth-law, as for class iii). This shows that some topological aspect of
the substrate -- namely the network on whose edges the spin variables are
defined --
is relevant in determining the 
growth-law and other universal properties. 
However this aspect cannot be traced back to usual indicators
such as the fractal or the spectral dimension. 

Despite that a precise identification of this topological feature is currently missing,
in \cite{noibarriers} the conjecture was proposed
that the same property might determine both the existence of the equilibrium 
phase-transition and the non-equilibrium asymptotic growth-law. Specifically, systems 
which do not sustain a (finite-temperature) para-ferromagnetic transition
(examples are the percolation network and the Sierpinski gasket) should belong to
class iii) and the others (i.e. the Sierpinski carpet) to class ii). 

Concerning the pre-asymptotic behavior, 
let us stress that, due to the recursive construction of the deterministic fractals discussed above,
the fractal properties are limited to distances larger than $\ell$, while for shorter
distances the structure is compact. This fixes the crossover length $\lambda =\ell$.
For $L(t)\ll \ell$ one recovers the characteristic growth-law (\ref{ldit}) of 
a homogeneous substrate.

\section{Coarsening in $d>1$}

Understanding coarsening in inhomogeneous systems in dimension 
larger than one is much harder than in $d=1$. 
On the one hand this is due to the fact that analytical approaches are
much more difficult, on the other hand because physical intuition
is less straightforward when interfaces are lines or surfaces that in the presence of the
pinning centers can bend and stretch. 
Moreover, numerical simulations - crucial in the absence of analytical tools - 
are very demanding due to the slow growth of $L(t)$.

Despite many similarities with the one-dimensional models, 
the pattern of behaviors exhibited in higher dimensions 
is different and in some cases quite enriched with respect to the $1d$ case, as we will discuss below.

\subsection{RFIM} \label{d2rf}

There is a general consensus on the fact that the RFIM is characterized by
a logarithmic growth both in $d=2$ \cite{rflog2,noirf,rao2} 
and in $d=3$ \cite{rflog3,noirf}, 
although the value of the exponent $\psi $ defined in Section \ref{downvsact} ii) 
cannot be precisely 
determined numerically, due to the very slow increase of $L(t)$.  
Recalling also the behavior
in $d=1$ this shows that the barriers encountered at late times have a similar nature in all
the spatial dimensions considered.
Experiments on real systems \cite{exprf} confirm the logarithmic growth.

The behavior of the model in the pre-asymptotic stage is, instead, more rich.
In \cite{rflog3} a pure-like
early-stage with Eq. (\ref{ldit}) was found. However in \cite{noirf} it was shown that,
by taking the small $T_f$ limit with fixed $h/T$, 
a different power-law, with an $h/T_f$ dependent exponent
$\zeta$ is observed, as in Eq. (\ref{plawT}). A similar behavior is observed in \cite{sanhi}. 

These apparently contrasting observations can be reconciled in a {\it double crossover}
scenario. With the two independent parameters $\epsilon =h/T_f$ and $\tau =T_f/J$ of the
model one can build two characteristic lengths $\lambda (\epsilon)$ and $\Lambda (\tau)$.
Then, in principle, one may observe a couple of crossovers as $L(t)$ meets these two
lengths, with different behaviors in the three time sectors separated by them
[furthermore, Eq. (\ref{scalg}) should be upgraded to contain the extra dependence of 
the scaling function $g$ on $\Lambda (\tau)/L(t_w)$]. 

In the low-$T_f$ limit considered in \cite{noirf} where $h/T_f$ is kept fixed, 
$\lambda $ is finite and $\tau \simeq 0$, implying also $\Lambda \simeq 0$.
This means that there is a single crossover 
at $L(t)\simeq \lambda $ from a pre-asymptotic to a late regime where 
the barriers seeded by the disorder act differently as to produce a change
from an initial homogeneous-like growth [Eq. (\ref{ldit})],
to the disorder-dependent power-law of Eq. (\ref{plawT}).
In the opposite situation of a large $T_f$ (actually as large as possible to keep
the ordering of the system) $\Lambda $ is rather large and $\lambda $ is so small
that the crossover associated to it, happening at very early times, is not observable.
One has again a single crossover, when $L(t)\simeq \Lambda$, but now it occurs
similarly to the $1d$ case, namely from a pre-asymptotic
behavior where barriers are tamed by thermal fluctuations, when one recovers Eq. (\ref{ldit}),
to the asymptotic logarithmic behavior.

In the articles cited above, the authors observe alternatively one or the other crossover,
and this is possibly due to the choice of the model parameters. If the {\it double crossover}
interpretation is correct it should in principle be possible, by tuning the parameters opportunely,
to observe the two crossovers in a single quench history. To the best of my
knowledge, however, this has not yet been reported. 

There is no consensus on the validity of superuniversality in this model. Indeed in
\cite{rflog3,otherfavorSU1} it was found to hold, while it was clearly shown not to be
obeyed in \cite{noirf}. However it must be recalled that the evidences in favor and against 
superuniversality are often obtained with different parameter settings.
For instance, the two situations with $T_f\simeq 0$ and $T_f$ finite are considered where,
as discussed above, also a different crossover pattern is observed.
This issue, therefore, is currently not well understood.

\subsection{SDIM} \label{sdim2}

Numerical simulations of this system in $d=2$ have been interpreted both in terms of 
a power-law increase of $L(t)$ \cite{otherfavorSU2,dil1}, or as a logarithmic
growth-law \cite{dil2}. A reconciliation between these interpretation is provided in 
\cite{noidil} in terms again of a rich crossover interplay, how it is briefly explained below.

\begin{figure}[h]
  \vspace{1cm}
    \centering
\rotatebox{0}{\resizebox{.75\textwidth}{!}{\includegraphics{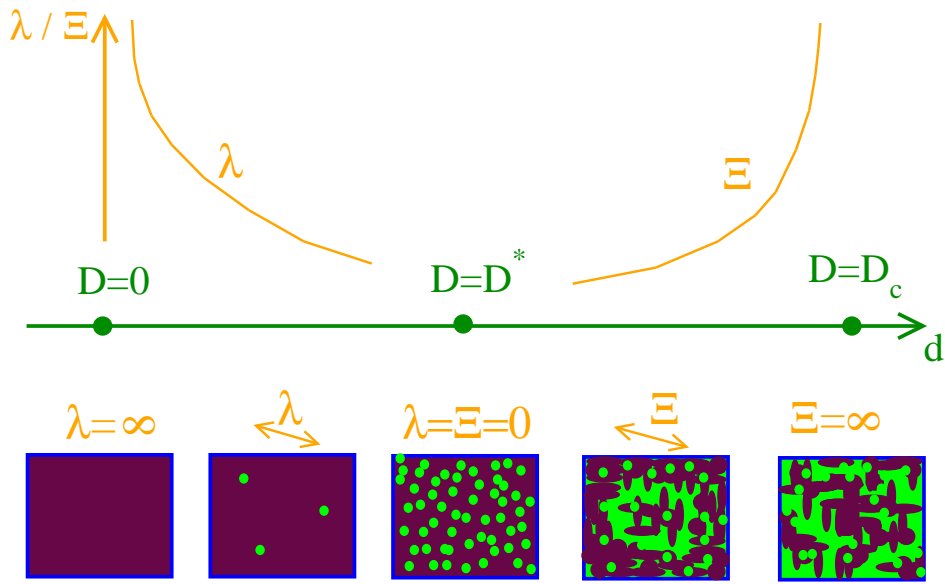}}}
\caption{A pictorial representation of the substrate (plum boxes below) as the
density $D$ of vacancies is changed (according to the green axis in the 
middle). The behavior of the two lengths $\lambda$ and $\Xi$ is shown in
orange.}
\label{substrate}
\end{figure}

Coarsening in this system is deeply related to the geometry of the diluted lattice.
Let us start, therefore, with a brief review of the geometrical properties of  
the substrate  
as dilution is varied from high to low values. According to percolation 
theory \cite{stauffer},
at $D = D_c$ a spanning fractal cluster of non-diluted sites is present, to which an infinite
percolative coherence length $\Xi (D_c)$ is associated (not to be confused with the
coherence length $\xi$ of the Ising model defined on top of the substrate).
This situation is pictorially represented on the far right of
figure \ref{substrate}.

As $D$ is slightly lowered below $D_c$ (i.e. moving a bit to the left
in figure \ref{substrate}),
the infinite cluster is fractal over distances up to 
$\Xi (D)\sim (D_c - D)^{-\nu}$, where $\nu =4/3$ is a critical exponent, and it becomes compact over 
larger distances. Besides, finite clusters are also present. 
Upon lowering further $D$, $\Xi $ keep decreasing until, for a certain value $D^*$ of $D$, it 
becomes comparable to the lattice spacing.
The behavior of $\Xi (D)$ is sketched as an orange 
line in the upper part of figure \ref{substrate} [i.e. for $D>D^*$].
For $D<D^*$, the infinite cluster is compact over all length scales and there are no finite droplets. There 
only remain isolated vacancies and, as already discussed in Sec. \ref{cross}, 
their typical distance defines another length
$\lambda (D)$ which keeps increasing upon lowering $D$ up to $\lambda(D=0)=\infty$, as shown in figure \ref{substrate} for $D<D^*$.

\begin{figure}[h]
  \vspace{1cm}
    \centering
\rotatebox{0}{\resizebox{.75\textwidth}{!}{\includegraphics{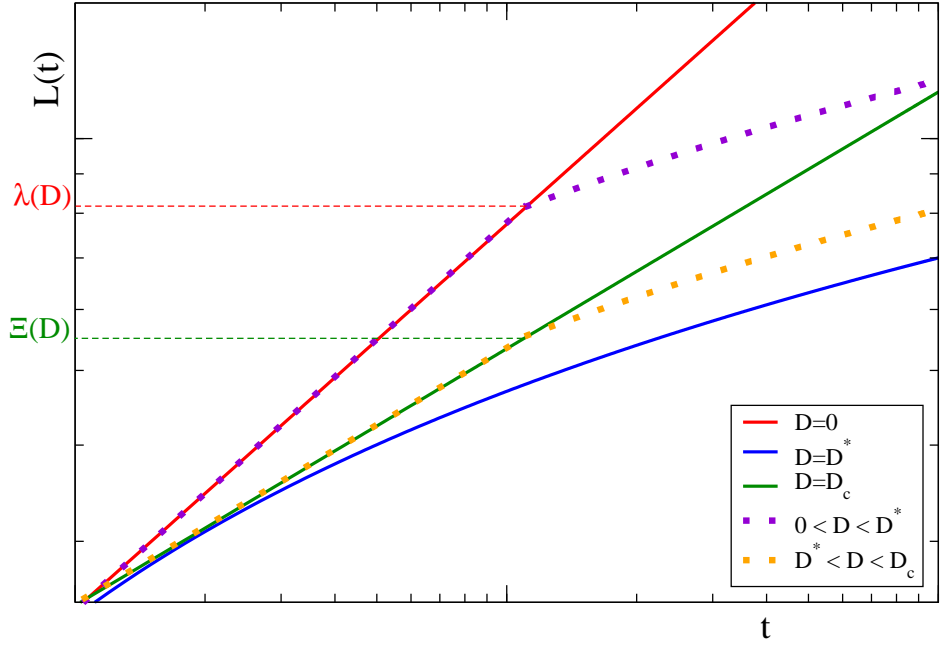}}}
\caption{A pictorial representation of the behavior of $L(t)$ for different
choices of $D$, on a double logarithmic scale.}
\label{figLdil}
\end{figure}

In the following we show how the conjecture put forth in \cite{noibarriers}, which was reviewed in Sec. \ref{fractals}, may help in understanding the kinetics of this model. 
Exactly at $D_c$, coarsening occurs on a percolation cluster. Since there is no ferromagnetic 
transition on this network, that is to say that $T_c=0$, 
a power-law growth as in Eq. (\ref{plawT}) is expected. 
This behavior is drawn in bold green in the schematic picture of 
figure \ref{figLdil}.
Away from the percolation point
$D=D_c$, the large-scale properties of the network provide an equilibrium 
phase-transition at a critical temperature $T_c(D)>0$: In this case, therefore, 
one expects an asymptotic logarithmic growth-law. This is recognized in 
figure \ref{figLdil} by the asymptotic downward bending of the curves
with $0<D<D_c$ for large times. Clearly, right at $D=0$ the homogeneous case
is recovered with the usual law (\ref{ldit}), which is shown in red in
figure \ref{figLdil}.
 
This pattern of behaviors is indeed observed 
in the numerical study \cite{noidil} of the model in the small $T_f$-limit. 
This supports the conjecture discussed in Section \ref{fractals}.

Not only the asymptotic laws, but also the early stage behaviors
can be understood. Let us start from the large dilution regime $D\lesssim D_c$.
In this case the properties of the spanning
cluster are the same as those at $D_c$ up the length $\Xi (D)$. Hence a crossover phenomenon
is expected from an early stage regulated by Eq. (\ref{plawT}) for $L(t)\ll \Xi (D)$ to
the asymptotic logarithmic growth when $L(t)\gg \Xi (D)$, as it is 
pictorially represented with a dotted-orange line in figure \ref{figLdil}. 
This shows that 
the growth-law may be interpreted as algebraic or logarithmic depending on the
choice of $D$ and on the time-scale of the observations and might explain why 
different findings are reported in the literature. 

Let us turn now to consider the small dilution regime, which behaves differently.
As discussed in Section \ref{cross}, a crossover occurs here  between an early homogeneous-like
behavior (\ref{ldit}) and a late logarithmic growth when $L(t)$ meets $\lambda (D)\sim D^{-1/d}$. This is shown in figure \ref{figLdil} with a
dotted-violet line. 

This whole pattern of crossovers is actually observed in \cite{noidil} suggesting - once again - 
that an interpretation in terms of the scaling symmetry  and
the related crossover structure is a groundbreaking tool. 

Notice also that, although two
lengths are present also in the SDIM, the crossover mechanism is very different from the one
- denoted as {\it double crossover} in Section \ref{d2rf} -
invoked for the RFIM. Indeed, in that case the two lengths $\lambda (\epsilon)$ and 
$\Lambda (\tau)$ are built from two independent parameters of the model. This allows
one, in principle, to have a double crossover, namely three consecutive regimes.

For the SDIM in the small-$T_f$ limit considered in \cite{noidil}, 
both $\Xi $ and $\lambda$ depend on the 
same parameter $D$ and this occur in such a way that either $\lambda \simeq 0$ 
(for $D>D^*$) or $\Lambda \simeq 0$ (for $D<D^*$). Hence, for any choice of $0<D<D_c$,
only one between the two lengths  $\lambda$ and $\Lambda$ can play a role, 
and a single crossover 
between an early and a late stage can be observed, as it is clarified 
in figure \ref{figLdil}.

Finally, let us consider the issue of superuniversality.
In Ref. \cite{dil2} \cite{dil1} the authors succeed in collapsing the curves for correlation functions
using the simple scaling (\ref{scalg}). In \cite{dil2}, this is interpreted as a possible confirmation 
of the superuniversality hypotheses. However in \cite{noidil} a study with a more 
ample choice of values of $D$ and different timescales clearly shows that (\ref{scalg})
is not sufficient to collapse all the curve and, instead, a scaling form with an extra 
argument as Eq. (\ref{scalgd}) is needed, suggesting
that superuniversality is violated also in this model. 
 
\subsection{RBIM} \label{rbim2}

For the RBIM, Huse and Henley (HH) \cite{HH}, elaborating on
previous ideas due to J. Villain \cite{villain}, predicted the logarithmic law (\ref{loglaw}) 
with $\psi=1/4$. Their argument, in $d=2$, can be summarized as follows: 
According to roughening theory a piece of interface of linear size $r$ 
in equilibrium deviates from
flatness of a typical quantity $w(r)\propto r^{\zeta_r}$, where the {\it roughening exponent}
$\zeta _r$ is a universal quantity. If disorder is present, the pinning energy of the 
interfaces scales as $E_p\sim r^\chi_r$, where $\chi_r$ is another exponent.

On the other hand, simply due to geometry, 
moving a distance $r$ along the surface of a coarsening 
curved domain one also deviates from flatness of a certain quantity 
$a(r,R)$, where $R$ is the radius of curvature. 
Hence, on scales $r$ up to $r_c$ - such that
$a(r,R)\simeq w(r)$ - the domain's wall is basically equilibrated.
A simple calculation shows that
$a(r,R)$ is at least of order $a=r^2/R$
(this can be very easily proved for a circular shape in the small $r$ limit) and hence
one has $r_c\simeq R^{1/(2-\zeta_r)}$.

The next smallest scale to evolve towards equilibrium is precisely $r_c$ and,
recalling the discussion above, the pinning energy associated to this is
$E_p\sim r_c^{\chi _r}\simeq R^\psi $, with $\psi={\chi_r/(2-\zeta _r)}$. 
For random bond disorder
in $d=2$ it is known \cite{HH,exprough} that $\zeta _r=2/3$ and $\chi _r=1/3$, leading to 
$\psi = 1/4$. Assuming scaling by the identification $R\sim L(t)$ and
plugging this result into ii) of Section \ref{downvsact} one arrives at
Eq. (\ref{loglaw}) with $\psi=1/4$.

The correctness of the HH prediction has remained controversial for quite a long
time since numerical results for the $d = 2$ RBIM showed a steady algebraic growth with a 
disorder-dependent dynamical exponent \cite{otherfavorSU1,henkelrb} as in 
Eq. (\ref{plawT}). This would imply an asymptotic logarithmic increase of the
pinning barriers at variance with the prediction of HH.
Recently, more extensive numerical simulations \cite{noirb} have produced
sufficient numerical evidence for the existence, after a long-lasting algebraic regime, 
of a crossover to an asymptotic logarithmic growth.
This supports the HH scenario, although a precise determination 
of the $\psi $ exponent is, also in this case, not possible.
Experiments on real two-dimensional systems also find \cite{exprb,exprb2,exprb3} 
a logarithmic growth and the HH exponent $\psi =4$ was reported in \cite{exprb2,exprb3}.

The very nature of the long-lasting pre-asymptotic regime with power-law increase
of the domain's size, which is observed also in some experiments \cite{exprb3},
 is not presently clarified. In view of the discussion of Section 
\ref{downvsact} it could be due to a genuine regime where the scaling of barriers
is only logarithmic -- according to iii) -- different therefore  from the asymptotic behavior
described by HH. This is not necessarily the case, since even within
the HH scenario long-lasting pre-asymptotic corrections might yield an effective
algebraic growth-law at intermediate times, as noticed in different contexts in
\cite{powervslog}.

Interestingly enough, the algebraic growth-law can be made truly asymptotic
by considering a symmetric bimodal distribution $J_{ij}=J_0\pm J_0$ of the coupling
constants, as it was shown numerically in \cite{noinuovo}. 
This choice amounts to have a fraction $1/2$ of 
bonds $J_{ij}=0$ and an equal quantity of $J_{ij}=2J_0$. This realizes a percolation
cluster of bonds, since the bond percolation threshold is $D_c=1/2$.
According to the conjecture mentioned in Sec. \ref{fractals}, one expects 
an algebraic increase of $L(t)$ as in Eq. (\ref{plawT}), as indeed it is observed. 

Regarding super universality, also in this model the situation is controversial:
while it was observed to hold true in \cite{otherfavorSU1,gyure},
clear violations  have been observed in \cite{noirb}.
Some experimental evidences in favor of a superuniversal scaling have been reported 
in \cite{exprb3}.

\section{Conclusions} 
 
In this Article I briefly reviewed the coarsening phenomena occurring 
in simple model-systems without space-translation invariance.
Following in the footsteps of early studies of phase-ordering in homogeneous 
ferromagnets, most efforts in this field have been devoted to the determination 
of the growth-law and of the dynamical scaling properties.
Apart from some instances of one-dimensional systems, 
analytical approaches are scarce and the current understanding 
of the problem rests mostly upon heuristic arguments and slowly
converging numerical simulations. This fact, together with the observation of 
a broad spectrum of distinct behaviors, make the problem far from being
settled.

Our poor understanding 
hinders the development of a  general framework 
for growth-kinetics in disordered and inhomogeneous ferromagnetic
systems, which is a subject of noteworthy thechnological relevance. 
This, in turn, encumbers the discussion on the very 
nature of more complex disordered systems, such as spin-glasses,
which according to a debated interpretation \cite{spglass} could also be 
described as a disordered ferromagnet.

As discussed in this paper, despite this state of affairs, the rich 
(and sometimes apparently contrasting) 
behavior of different systems -- or even of the same system 
in different temporal or parameter regions -- can be interpreted in terms
of an {\it upgraded} scaling framework where, next to the typical
size $L$ of ordered regions, there is another (or, in some cases, others)
relevant characteristic length $\lambda $ associated to the
inhomogeneities. This foster
the overview of the phenomenon and
could help to set the basis for a general renormalization group scheme
where diverse {\it universal } behaviors -- e.g. power-law
or logarithmic growth-laws -- are connected by a crossover pattern.  
However, the development of such a controlled theory
is matter for future research. 

\section*{Acknowledgments}

I acknowledge financial support by MURST PRIN 2010HXAW77 005. 

\vspace{.5cm}
Part of the project of this Special Issue on Coarsening took place at the Galileo Galilei Institute for Theoretical Physics in Arcetri, during the summer 2014 workshop {\it Advances in Nonequilibrium Statistical Mechanics}.

\end{document}